# Structural, elastic and electronic properties of Ir-based carbides-antiperovskites Ir$_3$$M$C ($M$ = Ti, Zr, Nb and Ta) as predicted from first-principles calculations

V. V. Bannikov[*], I. R. Shein, and D. V. Suetin

*Institute of Solid State Chemistry, Ural Branch of the Russian Academy of Sciences, 620990 Ekaterinburg, Russia*

Structural, elastic, electronic properties and the features of inter-atomic bonding in hypothetical Ir-based carbides-antiperovskites Ir$_3$$M$C ($M$ = Ti, Zr, Nb and Ta), as predicted from first-principles calculations, have been investigated for a first time. Their elastic constants, bulk, shear and Young's moduli, compressibility, Poisson's ratio, Debye temperature have been evaluated, and their stability, character of elastic anisotropy, brittle/ductile behavior, as well as electronic structure have been explored in comparison with binary carbides $M$C having NaCl-type structure. Authors hope that the presented results will be useful for future synthesis of these phases, as well as for extending the knowledge about the group of antiperovskite-type promising materials.

## 1. Introduction

The ternary carbides $M_3A$C (where $M$ are transition $d$-metals and $A$ are $sp$-elements of I-VI groups) with a cubic antiperovskite (or *inverse perovskite*) structure (space group *Pm-3m*) constitute a rather broad class of materials with a wide variety of promising physical properties. For instance, superconductivity near ferromagnetism with transition temperature $T_C \sim 8$ K was discovered for Ni$_3$MgC [1], and after that for a group of related Ni-based antiperovskite-type ternary carbides: Ni$_3$ZnC ($T_C \sim 2$ K) [2] and Ni$_3$CdC ($T_C \sim 2.5 - 3.2$K) [3], see also [4-5]. The interesting magnetocaloric effect was observed for a series of Mn- and Fe-based ternary carbides such as Mn$_3$GaC [6], Mn$_{3-x}$Fe$_x$SnC [7], Fe$_3$Zn$_{1-x}$Sn$_x$C [8], Fe$_3$AlC$_{1.1}$ [9], a spin glass-like behavior was established for Fe$_3$SnC [10], and the extremely low temperature coefficient of resistivity was found for Fe$_3$A$_x$Ga$_{1-x}$C (A = Cu, Ag) compositions [11]. Attention was paid also to experimental studies of structural, transport, mechanical, and thermodynamic properties of Mn$_3$AC (A = Al, Zn, Ga) [12-13], Ti$_3$AlC [14], Mn$_3$SnC [15], Ni$_3$In$_{0.95}$C [16], and some other antiperovskites. By now, a lot of first-principles band structure calculations has been performed to explore the structural, elastic, magnetic, and electronic properties of M$_3$AC phases, where M are Ti, Cr, Mn, Fe, Co, Ni, and A are Mg, Al, Ga, Sn, In, Tl, Cd, see [17-28] and References therein.

---

[*] Corresponding author.
  *E-mail address*: bannikov@ihim.uran.ru

Thus, as it is seen from this short overview of available publications, till now the main attention was paid to antiperovskite-type M$_3$AC ternary carbides, where M are 3d-metals, and A are sp-elements, whereas the experimental and theoretical data on the antiperovskites with M and A being the combinations of heavy 4d-, 5d-atoms and transition 3d-metals atoms - such as Co$_3$WC, Rh$_3$WC, Pt$_3$NbC, Pd$_3$ScC, are not numerous [29-34].

A decade ago, the authors of [31], using the empirical structural criteria of cubic antiperovskites stability (so-called tolerance factor t, indicating, if the sizes of octahedral voids in close-packed M$_3$A crystals (as the structural predecessors of M$_3$AC) are sufficient for embedding of B, C or N atoms, or they are not), have predicted a broad family of new ternary carbides M$_3$AC, where M are the atoms of heavy 4d- and 5d-metals, and A also are atoms of transition metals. In particular, it was proposed, that in the valid interval of tolerance factor (0.899 < t < 1.123) four Ir-based Ir$_3$MC carbides, namely Ir$_3$TiC, Ir$_3$ZrC, Ir$_3$NbC, and Ir$_3$TaC can exist. Certainly, this empirical approach does not provide any information about the physical properties of the predicted materials, simply being based on a choice of elements with suitable atomic radii. At the same time, the unusual electronic and magnetic properties may be expected for Ir$_3$MC due to interplay in antiperovskite structure of two sub-lattices formed with open-shell 5d-(3d,4d)-atoms, and the potential technological interest to Ir$_3$MC compounds may be assumed because of the presence in their structure of the platinum-group metal Ir (known as the material with high hardness) and carbon. Actually, the particular attention to the carbides and nitrides of platinum group metals (M = Ru, Rh, Pd, Os, Ir, Pt) has been arisen after the report of Ono *et al.* [35] about the successful synthesis of platinum monocarbide PtC at high *P-T* using the laser-heated diamond anvil cell technique, – especially as to materials with potentially superior hardness and minimal compressibility, see review [36]. The suppositions mentioned above follow from the search principle consisting in that the potential ultra-hard materials should combine the short and strong covalent bonds (due to the participation of 2*p*-atoms, such as B, C or N) with high valence charge density typical for the compounds of transition metals [37-39]. Besides, because of chemical stability and high melting point (~ 2454º C) of iridium, the Ir-based alloys and compounds draw attention as materials suitable for applications at very high temperatures.

Inspired by these reasons, we have investigated theoretically some properties (structural, elastic, electronic, and magnetic) for the series of hypothetical antiperovskite-like Ir-based ternary carbides Ir$_3$MC (M = Ti, Zr, Nb, and Ta) employing the first-principles FLAPW-GGA band method. Our data cover the optimized lattice parameters, formation energies, independent elastic constants, elastic moduli, sound velocities and Debye temperature, as well, as the Poisson's ratio, and Pugh's indicator of brittle/ductile behavior for the corresponding polycrystalline species. These obtained data are discussed in comparison with corresponding well-studied binary monocarbides *M*C of rock-salt type.

## 2. Models and computational aspects

The examined antiperovskites Ir$_3$MC have been considered in cubic structure (space group P*m-3m*, as supposed in [31]) consisting of *M* atoms at the corners, carbon at the body center, and Ir at the face centers of the cube. The atomic positions are Ir: 3*c* (½,½,0); *M*: 1*a* (0,0,0) and C: 1*b* (½,½,½).

Our calculations were carried out by means of the full-potential method with mixed basis APW+lo (LAPW) implemented in the WIEN2k suite of programs [40]. The generalized gradient approximation (GGA) of exchange-correlation potential in the PBE form [41] was used. Relativistic effects were taken into account within the scalar-relativistic approximation. The *muffin-tin* (MT) spheres radii were chosen to be 2.0 bohrs for Ir, Ti, Zr, Nb and 1.6 bohrs for C atoms. The energy *cut-off* separating atomic core and valence states was taken to be –8.0 Ry, the linearization energies $E_{l=2}$ for the atomic states of open *d*-shells were specified to be 0.7 Ry for Ir, 1.2 Ry for Ti, and 0.3 Ry for Zr and Nb. The plane-wave expansion was taken to $R_{MT} \times K_{MAX}$ equal to 7, and the ***k*** sampling with 10×10×10 *k*-points in the Brillouin zone (BZ) was used. The band structure calculations were performed for optimized values of lattice constants. The self-consistent computations were considered to be converged when the difference in the total energy of crystal (per unit cell) did not exceed 0.1 mRy and the difference in the electronic charge inside atomic MT spheres did not exceed 0.001 *e* as calculated at consecutive steps. To evaluate the possible magnetic effects, for all of Ir$_3$MC antiperovskites, the calculations were performed both in spin-restricted and spin-polarized (in assumption of the ferromagnetic spin ordering) variants. The hybridization effects were analyzed taking into account the densities of states (DOSs) obtained by a modified tetrahedron method [42]. Additionally, some peculiarities of inter-atomic bonding picture were visualized by means of charge density maps.

The values of three independent elastic constants ($C_{11}$, $C_{12}$ and $C_{44}$ in cubic symmetry) for Ir$_3$MC antiperovskites were estimated by applying the corresponding strains to initial cubic structure and calculating the energy of the distorted crystal. The following deformations were employed: hydrostatic compression (strain tensor $e_{xx}=e_{yy}=e_{zz}=\delta/3$, $e_{xy}=e_{xz}=e_{yz}=0$, where $\delta=V/V_0-1$), tetragonal volume-conserving distortion ($e_{xx}=e_{yy}= -e_t/3$, $e_{zz}=2\cdot e_t/3$, $e_{xy}=e_{xz}=e_{yz}=0$, where $e_t=c/a-1$), and rhombohedral distortion ($e_{xx}=e_{yy}=e_{zz}=\gamma/3$, $e_{xy}=e_{xz}=e_{yz}=2\cdot\gamma/3$, where $\gamma$ is relative variation of body diagonal) obtaining the values of $C_{11}+2\cdot C_{12}$, $C_{11}-C_{12}$, and $C_{11}+2\cdot C_{12}+4\cdot C_{44}$ linear combinations, respectively. Note at once, that the obtained values of $C_{ij}$ allow us to determine the mechanical stability of Ir$_3$MC using the well-known criteria for cubic crystals: $C_{44} > 0$, $C_{11}-C_{12} > 0$, and $C_{11}+2\cdot C_{12} > 0$. We have found that among four Ir$_3$MC three phases: Ir$_3$TiC, Ir$_3$ZrC, and Ir$_3$NbC satisfy this criteria, *i.e.* are intrinsically stable, whereas for Ir$_3$TaC the constant $C_{44}$ is negative. Therefore, further in our discussion we will focus on the aforementioned mechanically stable antiperovskites, for which the predicted properties will be discussed in comparison with the corresponding monocarbides TiC, ZrC, and NbC.

## 3. Results and discussion

*3.1. Structural parameters and formation energies.*

The optimized lattice constants ($a_0$) for mechanically-stable antiperovskites Ir$_3$MC (*M* = Ti, Zr, Nb) in comparison with binary monocarbides *M*C are listed in Table 1. While these values of $a_0$ predicted in [31] are practically identical (~4.01 Å), the calculated lattice parameters increase in the sequence $a_0$(Ir$_3$TiC)<$a_0$(Ir$_3$NbC)<$a_0$(Ir$_3$ZrC). This coincides with the same tendency for *M*C, *i.e.* the equilibrium volume of Ir$_3$MC antiperovskites is expected to be regulated by the kind of *M*C sub-lattice.

The inter-atomic distances Ir–C (2.047–2.082 Å) in Ir$_3$MC antiperovskites are close to the sum of covalent radii [43] of Ir and carbon atoms (~1.98 Å), this means that the directional Ir–C bonds in Ir$_3$MC should be most strong, whereas the directional C–M bonds ($d^{M-C}$ = 3.456–3.606 Å) should be practically absent. For Ir$_3$MC compounds the material density ($\rho$) evaluated directly is expected to be 2-3 times higher than that of corresponding MC carbides (Table 1), at the same time the predicted antiperovskites are less dense, than metallic fcc-Ir is ($\rho \sim$ 22.588 g/cm$^3$ [44]).

We have also estimated the formation energies (E$_{form}$) of Ir$_3$MC assuming the one of their possible synthesis ways - from the metallic iridium and corresponding monocarbides, *i.e.*, with formal reaction: 3(fcc-Ir) +MC → Ir$_3$MC as: E$_{form}$ = E$_{tot}$(Ir$_3$MC) – [3E$_{tot}$(fcc-Ir) + E$_{tot}$(MC)], where E$_{tot}$ are the calculated total energies (per formula unit) of corresponding substances. Within this definition, a negative E$_{form}$ value would indicate the phase stability of the reaction product with respect to the solid-state mixture of initial reagents, and *vice versa*, if E$_{form}$ is positive. According to our calculations, for all modeled Ir$_3$MC antiperovskites the values of E$_{form}$ are found to be positive: +1.209 eV for Ir$_3$TiC, +0.610 eV for Ir$_3$ZrC, and +1.950 eV for Ir$_3$NbC, so the supposed synthesis from stable fcc-Ir and MC carbides (with negative E$_{form}$ with respect to corresponding elementary substances [45]) is expected to be endothermal reaction with very large enthalpy $\Delta H$>0, which definitely is improbable at the ambient conditions. Nevertheless, the majority of carbides and nitrides of platinum-group metals examined theoretically by now, also adopt E$_{form}$>0 of the same order of magnitude, and their synthesis by the conventional methods seems to be very problematic, on the other hand, some of them have been successfully synthesized using a high pressure - high temperature technique – PtC [35] and Ru$_2$C [46], for instance, or reactive pulse laser deposition – RuN [47], as an example. So we hope, that Ir$_3$MC antiperovskites under consideration also can be synthesized employing these advanced methods.

*3.2. Elastic properties.*

At first, let us discuss the elastic properties of Ir$_3$MC *monocrystals* in comparison with those of corresponding MC monocarbides. The calculated elastic constants $C_{ij}$ for these compounds are listed in Table 1, for MC monocarbides these values are in reasonable agreement with available data (see [45,48-49] and references therein), for Ir$_3$MC antiperovskites they were obtained for a first time (for mechanically unstable Ir$_3$TaC the $C_{ij}$ constants were calculated to be $C_{11}$ = 370.4 GPa, $C_{12}$ = 252.1 GPa, and $C_{44}$ = –43.4 GPa). The estimated values of bulk modulus $B_0=(C_{11}+2·C_{12})/3$, compressibility $\beta=1/B_0$, tetragonal shear modulus $G_t=(C_{11}–C_{12})/2$ and Cauchy pressure $CP=(C_{12}–C_{44})$ also are collected there, and the Blackman`s diagram for the series of considered antiperovskites and monocarbides is depicted in Fig.1, for obviousness, the metallic *fcc*-Ir is also shown there, the data for it are taken from [50]. As is seen, in contrast with monocarbides and metallic Ir characterized with negative Cauchy pressure indicating the predominant role of covalent bonds [51], the comparatively large positive values of *CP* and, as a consequence, the prevalence of metallic chemical bonding are expected for Ir$_3$MC antiperovskites. It is known, that the dependence of Young`s modulus (*Y*) on the selected direction ($\boldsymbol{n}=(n_x, n_y, n_z)$, $|\boldsymbol{n}|=1$) for a cubic crystal is given by:

$$Y(\boldsymbol{n}) = [(C_{11}+C_{12})/\{(C_{11}+2·C_{12})·(C_{11}–C_{12})\} + \\ + (1–A_Z)·(n_x^2·n_y^2+ n_x^2·n_z^2+ n_y^2·n_z^2)/C_{44}]^{-1} \quad (1)$$

where $A_Z = 2·C_{44}/(C_{11} − C_{12})$ is Zener anisotropy index, if $A_Z=1$, *Y* does not depend on the direction $\boldsymbol{n}$, while if $A_Z>1$, the Young`s modulus is maximal in [111] and minimal in [100] direction, and *vica versa*, if $A_Z<1$; the greater

|1–$A_Z$| is, the more considerable the "dispersion" of $Y$ becomes. It is seen from Table 1 that both for Ir$_3$$M$C and $M$C carbides $A_Z$<1, so the maximal value of Young`s modulus for them corresponds to the fourfold axes of monocrystals, and minimal one – to [111] directions. For all Ir$_3$$M$C antiperovskites the deviation of $A_Z$ value from unity is greater than it is for respective monocarbides, so for the formers the $Y$ magnitude is expected to be more sensitive to the choice of direction in crystal (see the $Y$ values calculated for [100], [110] and [111] directions in Table 1). At the same time, it is seen in Fig.2, where the broaching of Young`s modulus dependence on the selected direction in (110) plane is shown both for Ir$_3$$M$C and $M$C monocrystals, that the character of $Y$ "dispersion" strongly depends also on the type of $M$ atom. Actually, if $M$=Zr, the relatively weak dependence of $Y$ on the direction occurs – both for Ir$_3$ZrC and ZrC, however, if $M$=Nb, the Young`s modulus "dispersion" is maximally sharp. Note, that the variation of Young`s modulus for NbC monocarbide is considerably non-monotonic (the difference between its local maximum and minimum in [110] and [111] directions, respectively, is about 35 GPa), whereas for Ir$_3$NbC monocarbide the magnitude of $Y$ harshly decreases (~140 GPa) going from [100] to [111] direction, varying just slightly (~7 GPa) between [111] and [110] directions. For monocarbides the maximal value of $Y$ is considerably (~2–3.5 times) higher than for corresponding antiperovskites. In turn, the Poisson`s ratio value calculated as:

$$\nu(\boldsymbol{n})=[1-Y(\boldsymbol{n})/(3 \cdot B_0)]/2, \quad (2)$$

for Ir$_3$$M$C in all specified directions [100], [110] and [111] is greater than for $M$C carbides, and its largest magnitude $\nu$~0.48, close to the limit value of 0.5, is expected for [111] direction in Ir$_3$NbC.

Since the most of compositions are synthesized and utilized as the ceramic materials, not the monocrystals, the estimation of elastic characteristics of *polycrystalline* samples seems to be in interest. Those can de evaluated, for instance, in Voigt-Reuss-Hill (VRH) approximation [52] based on the values of $C_{ij}$ elastic constants for monocrystals obtained before. For cubic Bravais lattice, the polycrystalline bulk modulus both in Voigt and Reuss approaches equals to that of monocrystals, the polycrystalline shear modulus in Voigt and Reuss approximations is

$$G_V=(C_{11}-C_{12}+3 \cdot C_{44})/5 \text{ and } G_R=5 \cdot (C_{11}-C_{12})/[4+3 \cdot (C_{11}-C_{12})/C_{44}], \quad (3)$$

respectively, and finally, in VRH approach it is taken as arithmetic mean $G_{VRH} = (G_V+G_R)/2$. The results are listed in Table 2. As is seen, the values of polycrystalline bulk modulus for Ir$_3$MC antiperovskites and MC monocarbides are comparable (~220-300 GPa), whereas the values of shear modulus $G$ for Ir$_3$MC are considerably less than those are for MC (so, for Ir$_3$NbC the value of $G$ is order of magnitude less than for NbC), as well, as the values of isotropic Young`s modulus $Y_{VRH}=(9 \cdot B_0)/(1+3 \cdot B/G_{VRH})$ are (for Ir$_3$NbC it is about 7.5 times less as compared with NbC). Hence, it may be expected that the Ir-containing antiperovskites under consideration should be less stiff with respect to low-symmetry deformations (such as shears, monoaxial extension, etc.), than the corresponding monocarbides MC are. The variation of polycrystalline bulk, shear and Young`s moduli is visualized in Fig.3, as going from MC to Ir$_3$MC. According to the Pugh`s criteria [53], the material behaves in a brittle manner, if $G/B>0.571$, and is ductile, if inversely. As is seen from Table 2 and Fig.4, while both all the MC monocarbides and metallic iridium are brittle, the Ir$_3$MC antiperovskites are expected to be ductile, being relatively far enough from the "ductility-brittleness" border. The obtained results agree with the above conclusion on predominating metallic contribution to the chemical bonding in Ir$_3$MC antiperovskites, as compared with MC monocarbides, where the covalent bonding component prevails.

As the elastic characteristics of polycrystalline species (treated as isotropic substance) have been estimated, it is possible to evaluate the velocities of transverse ($v_\perp$) and longitudinal ($v_\parallel$) sound waves in the material as:

$$v_\perp = [G/\rho]^{1/2}, \text{ and } v_\parallel = [(B+4 \cdot G/3)/\rho]^{1/2} \quad (4).$$

It is seen (Table 2), that both $v_\parallel$ and $v_\perp$ in Ir$_3$MC species are predicted to be lower as compared with corresponding MC carbides (about 2 and 3-4 times, respectively). Further, there is a conventional way [45, 48, 54] to estimate the Debye temperature ($T_D$) of polycrystalline solids as follows:

$$T_D = (h \cdot v_s / k_B) \cdot [3s \cdot N_A \cdot \rho/(4\pi \cdot M)]^{1/3} \quad (5),$$

where $h$, $k_B$ and $N_A$ are Planck, Boltzmann and Avogadro constants, respectively, $M$ is the molecular weight of compound formula unit, $s$ is the number of atoms per one, and $v_s$ is the averaged sound velocity defined as:

$$(3/v_s^3) = (1/v_\parallel^3) + (2/v_\perp^3) \quad (6).$$

The expression $[\ldots]^{1/3}$ in (5) represents the reciprocal value of averaged inter-atomic distance in solid taken as a measure of minimal length of sound wave able to propagate through the crystal. Though the universal validity of this rough approach is undecided, it seems to be more or less reasonable as applied to close-packed crystals. At least, for TiC and NbC it provides the values of $T_D$ being in satisfactory agreement with available experimental data (see [48], Table 4 and references therein), and $T_D$ values for MC monocarbides estimated within the structural and elastic parameters reported here also are in reasonable accordance with the results obtained before (Table 2). As for Ir$_3$MC antiperovskites, the calculated $T_D$ values for them are 285 K, 218 K and 166 K for $M$ = Ti, Zr, Nb, respectively, being 3-4 times less, as compared with the corresponding MC monocarbides. The temperature dependence of lattice vibrational contribution to the molar heat capacity of solids in the Debye model can be computed straightforward, as follows:

$$C_V(T) = s \cdot 3R \left\{ 12\left(\frac{T}{T_D}\right)^3 \int_0^{T_D/T} \frac{x^3}{\exp(x)-1} dx - \frac{3T_D/T}{\exp(T_D/T)-1} \right\} \quad (7),$$

where $R$ is the universal gas constant (note, that formula (7) provides $C_V$ value related to the mole of *compound*, while the standard Debye theory deals with the mole of *atomic oscillators* (i.e., atoms of *any* sorts) in solid, so, one mole of $Ir_3TiC$, for instance, is treated as five moles of oscillators). As is seen from Fig.5, where the obtained $C_V(T)$ curves are plotted, the molar heat capacity of $Ir_3MC$ antiperovskites at low temperatures (being proportional to $(T/T_D)^3$) grows with $T$ considerably more steeply and approximates to the classical Dulong-Petit limit at significantly lesser $T$ values, as compared with $C_V$ of $MC$ monocarbides. From the other hand, the dependence of $C_V$ on $T$ in the vicinity of room temperature (~300 K) for antiperovskites is expected to be weaker than it takes place for monocarbides (see Inset in Fig.5, where behavior of the derivative $(dC_V/dT)$ *vs* $T$ is shown). In principle, taking into account also the electronic contribution to heat capacity (see below), it would be possible to estimate and compare other thermodynamic properties both of $Ir_3MC$ and $MC$ compounds (as far, as rough Debye model was supposed to be sufficient). However, we have decided not to go beyond the results discussed above, because the purpose of this paper is only the initial estimation of $Ir_3MC$ properties with the hope to motivate the interest to this new sub-group of antiperovskites.

*3.3. Band structure, electronic and magnetic properties*

Now we shall discuss in brief the features of band structure and related electronic and magnetic properties of $Ir_3MC$ antiperovskites in comparison with those of $MC$ monocarbides. As is seen in Figs. 6-7, where the calculated total and partial DOSs are depicted, both the predicted $Ir_3MC$ and *M*C compounds are expected to possess metallic conductivity, however, the DOS at Fermi level, N($E_F$), and in its vicinity for antiperovskites is an order of magnitude higher than for corresponding monocarbides (Table 3). So, it can be anticipated that predicted Ir-containing antiperovskites would be characterized with considerably greater concentration of conduction electrons and, as a consequence, with higher related transport properties (such as electrical and thermal conductivity), as well, as with higher electronic contribution to their heat capacity ($C_V^{(el)}=\gamma \cdot T$, where $\gamma$ is the Sommerfeld constant equal to $(\pi \cdot k_B)^2 \cdot [N_\uparrow(E_F)+N_\downarrow(E_F)]/3$, see Table 3) than the corresponding *M*C monocarbides are.

While the electronic spectra of *M*C monocarbides are quite simple, being composed with hybridized *M-d* and C-2*p* states and characterized with sharp DOS peak in the middle of the valence band and low-density plateau in the near-Fermi region (shifting towards lower energies with respect to $E_F$, as going from TiC and ZrC to NbC, in accordance with rigid band model), for Ir$_3$*M*C antiperovskites the distribution of valence states density is much more complex (Figs. 6-7). First, for Ir$_3$*M*C the valence band is formed predominantly with Ir-5*d* states with only small admixture of M-*d* and C-2*p* states (responsible for the covalent contribution to the chemical bonding, see below) in its middle and in the vicinity of $E_F$ (note, that there is a gap for M-*d* and C-2*p* states in antiperovskites). Next, the DOS distribution for Ir$_3$*M*C adopts the complicated multi-peak structure, which may be explained quantitatively as follows: the symmetry of surrounding of Ir atoms in antiperovskite cell is tetragonal, hence the Ir-5*d* band splits into t$_{2g}$(*xz, yz*), t$_{2g}$(*xy*), e$_g$($z^2$) and e$_g$($x^2-y^2$) sub-bands resulting in the complex form of DOS profile.

It is interesting to note, that for Ir$_3$TiC the near-Fermi Ti-3*d*$_{\uparrow,\downarrow}$ electronic states suffer considerable spin polarization resulting in magnetic moments (~0.61 $\mu_B$) induced on Ti atoms and, as a consequence, in magnetic ground state of Ti-containing antiperovskite (the calculated total magnetic moment per *equilibrium* unit cell ($\mu_0$) is about 0.907 $\mu_B$, the ferromagnetic ordering was supposed in modeling) – unlike non-magnetic TiC binary carbide. This result is quite unexpected, because in the majority of known Ti-containing crystalline substances the titanium atoms are present in non-magnetic state. As for $Ir_3ZrC$ and $Ir_3NbC$, they were predicted to be non-magnetic (the spin polarization of the valence band is negligible), as well, as the corresponding monocarbides are. It is known, that the simplest theory of ferromagnetism of collectivized conduction electrons results in Stoner criteria, alleging that ferromagnetic state is stable, if $I \cdot N(E_F) > 1$ ($I = 2\mu_B^2 \cdot \lambda$, where $\lambda$ is the constant of Weiss molecular field). This criteria naturally clarifies the predicted stabilization of $Ir_3TiC$ magnetic state simply with increase of $N(E_F)$ as compared with TiC, however, it cannot explain why the ground state of Zr- and Nb-based antiperovskites remains non-magnetic, despite high growth of $N(E_F)$ with respect to corresponding monocarbides. At the same time, the Stoner criteria implies the smooth behavior of DOS in the vicinity of $E_F$ (like $N(E) \sim E^{1/2}$, as it follows from the standard theory of free electron gas) and it may fail if the $N(E)$ dependence is more complex – sharp growth or presence of extremum near $E_F$, for instance, as it takes place for $Ir_3NbC$ and $Ir_3ZrC$, respectively (Fig.7). But the detailed consideration of this problem is beyond this paper confined only with preliminary modeling of properties of Ir-containing antiperovskites. Further, it was found for $Ir_3TiC$, that the total magnetic moment per unit cell ($\mu$) varies with mechanical strains being applied to the crystal. In Fig. 8 the estimated variation of $\mu/\mu_0$ ratio both with "hydrostatic" expansion (compression) and volume-conserving tetragonal distortion is shown (the strain parameter $\Delta$ was defined as $V/V_0 -$ 1 and $c/a-1$, respectively). It is seen that the reduction of $\mu/\mu_0$ magnitude may achieve ~8-12% at strains $|\Delta|\leq 0.04$, so some pressure-induced magnetic effects (like the Villari effect) may be expected in Ir$_3$TiC. We have estimated also the contribution of conduction electrons into the magnetic susceptibility of compounds under consideration (or Pauli paramagnetic susceptibility) as $\chi_P=[N\uparrow(E_F)+N\downarrow(E_F)]\cdot\mu_B^2$ (see Table 3), which is simply proportional to $N(E_F)$, so for Ir$_3$MC antiperovskites $\chi_P$ is expected to be order of magnitude higher than it is for corresponding MC monocarbides. For Ir$_3$TiC, though, the others contributions to magnetic susceptibility associated with non-zero spontaneous magnetization are supposed to be predominant.

*3.4. Chemical bonding*

As a final issue, let us examine some features of chemical bonding picture in Ir$_3$MC antiperovskites in comparison with MC monocarbides. We shall do that regarding only the pair Ir$_3$ZrC *vs* ZrC as an example, because the bonding picture in antiperovskites with M=Ti, Zr, Nb appears to be quite similar (except only the aspect that there is non-zero differential spin density $\sigma\uparrow-\sigma\downarrow$ on Ti atoms in magnetic Ir$_3$TiC, however it does not change drastically the general characteristics of bonding). In the Fig.9 the calculated maps of valence charge density distribution in specified crystallographic planes of ZrC and Ir$_3$ZrC are shown. It is seen, that in ZrC binary carbide the bonding picture is quite simple: the distinctive directed bonds take place only between nearest Zr and C atoms assuming the considerable role of covalent contribution, meanwhile, the effective atomic charges calculated in the frameworks of Bader`s approach [55], $Q$(Zr) and $Q$(C) were estimated to be ±1.88, respectively (Table 3), however, the "metallic" contribution to chemical bonding picture due to the conduction electrons is supposed to be insignificant – because of their small concentration (see *Sec. 3.3*). So the chemical bonding of complex covalent-ionic character takes place in ZrC, as is common for the classical "metal cation - non-metal anion" binary compound, and the compound should behave in a brittle matter – in accordance with the results discussed before (see *Sec. 3.2*).

For Ir$_3$ZrC antiperovskite the bonding picture is much more complex and intriguing. First, as is seen in Fig.9-b, the additional directed Zr-Ir bonds are formed - due to the slight hybridization of Zr-4$d$ and Ir-5$d$ electronic states in the middle of the valence band about 2.5 eV below E$_F$, see Fig.7, where their overlapping is shown by arrow (at the same time, these states overlapped in the vicinity of E$_F$ are supposed to be non-bonding). Note, that the similar formation of directed "metal-metal" bonds also takes place in metallic *fcc*-Ir and in some other inter-metallic compositions. Next, the directed Zr-C bonds are absent in Ir$_3$ZrC antiperovskite (due to the increased inter-atomic distance), on the other hand, the sharply distinctive Ir-C bonds characterized with relatively high valence charge density in corresponding direction have been formed (Fig.9.c) as a result of Ir-5$d$ and C-2$p$ states hybridization about 4-6 eV below E$_F$ (Fig.7). So, the role of the covalent contribution to the chemical bonding in Ir$_3$ZrC still seems to be essential, and this, at first sight, is in contradiction with earlier result on its ductile behavior typical for the compounds with predominant metallic character of bonding. However, an intriguing point should be taking into account here. As the effective atomic charges in Ir$_3$ZrC have been calculated (Table 3), the quite unexpected result has been obtained, namely, the *negative* effective charge value on Ir atoms, $Q(\text{Ir})= -0.33$ $e$. It should be remembered, though, that the Bader`s approach is based simply on the analysis of total charge density distribution (searching for its minima surfaces), so no evident reason to interpret this result as "an anionic state" of metal atoms in antiperovskite, it simply means the excess electronic density presence in their neighborhood. Since the $Q$(C) value is also negative (–0.80 $e$), and $Q$(Zr) is positive (+1.79 $e$), the following model of chemical bonding in Ir$_3$ZrC can be proposed, in accordance with these results and Fig.9.c. There is a simple cubic sub-lattice of positively charged Zr$^{+\delta}$ ions with the network of negatively charged [Ir$_6$C]$^{-\delta}$ octahedra implanted it its voids and regarded as unified charged centers with no strict separation of electronic charge between iridium and carbon atoms, and this [Ir$_6$C] network is coupled with Zr sub-lattice both by interaction of ionic character and directed metal-metal bonds. As is seen in Fig.9.c, the electronic charge density in Ir$_3$ZrC is distributed mainly inside [Ir$_6$C] octahedra both around C and Ir atoms, that may be interpreted as the increase of the role of collectivized electrons in the specification of the chemical bonding character. So, besides the covalent component of bonding, the role of "metallic" contribution of conduction electrons also is expected to be important, specifying the predicted metallic-like properties of Ir$_3$ZrC, such as its ductility. Note, that for Ir$_3$TiC and Ir$_3$NbC the similar situation with calculated effective atomic charges takes place, so the general picture of chemical bonding in the family of Ir-based antiperovskites is expected to be also similar to that, described above.

## 4. The Conclusion

In summary, the theoretical comparative study of structural, elastic, electronic, magnetic properties, band structure and chemical bonding picture of the hypothetical Ir$_3$$M$C antiperovskites *vs* $M$C monocarbides ($M$ = Ti, Zr, Nb) have been performed for a first time using the first-principles FLAPW-GGA method as the computational implement. It was predicted that Ir$_3$$M$C antiperovskites are metastable with respect to the solid-state mixture of metallic *fcc*-Ir and corresponding $M$C monocarbides with mixing enthalpy $\Delta H$~1-2 eV, so their formation at the ambient conditions seems to be improbable, and the synthesis techniques at non-equilibrium conditions (like reactive pulse laser deposition) should be employed. The monocrystals of Ir$_3$MC antiperovskites are characterized with Young`s modulus being more sensitive to the direction selected in crystal and are less stiffer with respect to low-symmetry deformations (such as shear) than MC monocarbides are. While all the MC monocarbides under consideration are brittle, for Ir$_3$MC the ductile behavior is expected. The estimated Debye temperature values for antiperovskites appear to be 3-4 times less than for corresponding monocarbides. According to the results of band structure calculations, all the Ir$_3$MC antiperovskites are expected to exhibit metallic conductivity, moreover, the concentration of conduction electrons for them is expected to be much higher than for MC monocarbides, and as a consequence, for the formers the better transport properties are expected. Ir$_3$TiC is predicted to be magnetic compound with magnetization being slightly sensitive to the external mechanical strains, while Ir$_3$ZrC and Ir$_3$NbC, as well, as all the MC monocarbides, are non-magnetic metals. In contrast to MC monocarbides, where the role of covalent-ionic contribution to the chemical bonding is predominant, in Ir$_3$MC antiperovskites besides the covalent component of bonding, the role of "metallic" contribution also is expected to be important, and this results in the coexistence of well resolved directed inter-atomic bonds and ductile behavior of these materials to be predicted.

**Table 1**

Calculated lattice constants ($a_0$, in Å), material density ($\rho$, in g/cm$^3$) and elastic parameters for antiperovskite-like ternary carbides Ir$_3$MC (M = Ti, Zr, Nb) *monocrystals* in comparison with monocarbides MC: independent elastic constants ($C_{ij}$, in GPa), bulk moduli ($B_0$, in GPa), compressibility ($\beta$, in GPa$^{-1}$), tetragonal shear moduli ($G_t$, in GPa), Cauchy pressure (CP, in GPa), Zener index of elastic anisotropy ($A_Z$), Young's moduli and Poisson's ratios in the [100], [110] and [111] directions ($Y_{[100],[110],[111]}$, in GPa; $\nu_{[100],[110],[111]}$).

|  | TiC | ZrC | NbC | Ir$_3$TiC | Ir$_3$ZrC | Ir$_3$NbC |
|---|---|---|---|---|---|---|
| $a_0$ | 4.337 (4.27/4.42)[a] | 4.710 (4.69/4.71)[a] | 4.487 (4.43/4.49)[a] | 4.095 | 4.164 | 4.122 |
| $\rho$ | 4.876 | 6.563 | 7.714 | 15.393 | 15.637 | 16.160 |
| $C_{11}$ | 518.4 (467/603)[a] | 456.9 (446/470)[a] | 654.8 (557/646)[a] | 380.4 | 325.3 | 366.6 |
| $C_{12}$ | 123.7 (97/123)[a] | 107.2 (100/103)[a] | 125.3 (121/200)[a] | 208.0 | 237.1 | 246.3 |
| $C_{44}$ | 162.9 (129/206)[a] | 148.9 (138/160)[a] | 170.2 (142/192)[a] | 51.9 | 35.1 | 9.6 |
| $B_0$ | 255.3 | 223.7 | 301.8 | 265.5 | 266.5 | 286.4 |
| $\beta$ | 0.0039 | 0.0045 | 0.0033 | 0.0038 | 0.0036 | 0.0035 |
| $G_t$ | 197.4 | 174.6 | 264.7 | 86.2 | 44.1 | 60.3 |
| CP | -39.2 | -41.7 | -44.9 | 156.1 | 202.0 | 236.7 |
| $A_Z$ | 0.8254 | 0.8516 | 0.6429 | 0.6021 | 0.7959 | 0.1596 |
| $Y_{[100]}$ | 470.7 | 416.2 | 614.5 | 233.3 | 125.4 | 168.6 |
| $Y_{[110]}$ | 418.0 | 377.1 | 464.7 | 161.2 | 106.1 | 35.9 |
| $Y_{[111]}$ | 403.0 | 365.6 | 429.8 | 146.2 | 100.9 | 28.5 |
| $\nu_{[100]}$ | 0.1926 | 0.1900 | 0.1606 | 0.3535 | 0.4216 | 0.4019 |
| $\nu_{[110]}$ | 0.2271 | 0.2191 | 0.2434 | 0.3988 | 0.4337 | 0.4791 |
| $\nu_{[111]}$ | 0.2369 | 0.2277 | 0.2626 | 0.4082 | 0.4369 | 0.4834 |

[a] The minimal/maximal values of available data, as taken from [45, 48-49] (see also References therein).

**Table 2**

Calculated elastic parameters for *polycrystalline* antiperovskites Ir$_3$MC (M = Ti, Zr, Nb) in comparison with those of MC monocarbides: bulk moduli ($B_V=B_R=B$, in GPa), shear moduli ($G_V$, $G_R$, $G$, in GPa), isotropic Young's moduli ($Y$, in GPa), Poisson's ratio ($v$), and Pugh's indicator ($G/B$). The estimated velocities of transverse ($v_\perp$), longitudinal ($v_\parallel$) and averaged ($v_s$) sound waves (in 10$^3$ m/s), as well, as the Debye temperature values ($T_D$, in K) also are listed here.

| Parameters | TiC | ZrC | NbC | Ir$_3$TiC | Ir$_3$ZrC | Ir$_3$NbC |
|---|---|---|---|---|---|---|
| $B_V = B_R = B$ | 255.3 | 223.8 | 301.8 | 265.5 | 266.5 | 286.4 |
| $G_V$ | 176.7 | 159.3 | 208.0 | 65.6 | 38.7 | 29.8 |
| $G_R$ | 175.1 | 158.3 | 301.8 | 61.7 | 38.2 | 14.5 |
| $G = (G_V + G_R)/2$ | 175.9 | 158.8 | 203.3 | 63.7 | 38.5 | 22.1 |
| $Y$ | 429.1 | 385.2 | 498.0 | 176.9 | 110.1 | 64.8 |
| $v$ | 0.2198 | 0.2131 | 0.2249 | 0.3889 | 0.4312 | 0.4623 |
| $G/B$ | 0.6891 | 0.7096 | 0.6736 | 0.2398 | 0.1443 | 0.0773 |
| $v_\perp$ | 6.01 | 4.92 | 5.13 | 2.03 | 1.57 | 1.17 |
| $v_\parallel$ | 10.02 | 8.15 | 8.62 | 4.77 | 4.51 | 4.42 |
| $v_s$ | 6.65 | 5.44 | 5.68 | 2.29 | 1.78 | 1.34 |
| $T_D$ | 913 (981/921)* | 688 (670/-) | 754 (681/739) | 285 | 218 | 166 |

*The calculated data taken from [45]/[48], respectively, are specified in parenthesis.

**Table 3.** Calculated spin-polarized DOSs at $E_F$ ($N_{\uparrow,\downarrow}(E_F)$, in states/eV, per formula unit), Sommerfeld constant ($\gamma$, in mJ·K$^{-2}$·mole$^{-1}$), Pauli paramagnetic susceptibility ($\chi_P$, in 10$^{-4}$ cm$^3$·mole$^{-1}$) and effective atomic charges ($Q(M)$, $Q(C)$, $Q(Ir)$, in $e$ units) for Ir$_3$MC antiperovskites (M = Ti, Zr, Nb) in comparison with corresponding monocarbides MC.

| Parameters | TiC | ZrC | NbC | Ir$_3$TiC | Ir$_3$ZrC | Ir$_3$NbC |
|---|---|---|---|---|---|---|
| $N_\uparrow(E_F)$ | 0.1125 | 0.1860 | 0.7309 | 1.6220 | 2.6390 | 3.2323 |
| $N_\downarrow(E_F)$ | 0.1125 | 0.1860 | 0.7309 | 1.3948 | 2.6353 | 3.2448 |
| $\gamma$ | 0.5305 | 0.8771 | 3.4468 | 7.1133 | 12.4363 | 15.2723 |
| $\chi_P$ | 0.07 | 0.12 | 0.47 | 0.98 | 1.71 | 2.10 |
| $Q(M)$ | +1.68 | +1.88 | +1.74 | +1.50 | +1.79 | +1.39 |
| $Q(C)$ | -1.68 | -1.88 | -1.74 | -0.81 | -0.80 | -0.82 |
| $Q(Ir)$ | - | - | - | -0.23 | -0.33 | -0.19 |

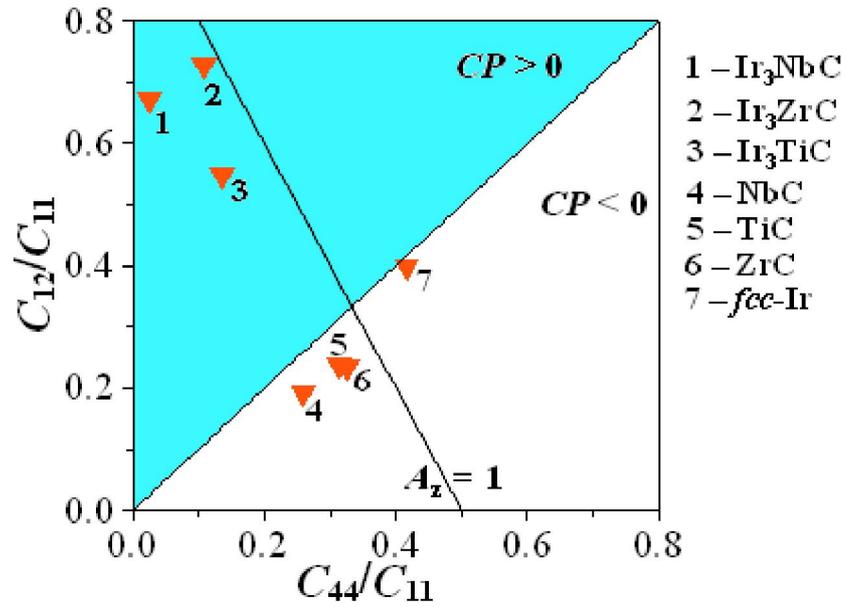

**Fig. 1.** The Blackman's diagram for $Ir_3MC$ antiperovskites and $MC$ monocarbides series ($M$ = Ti, Zr, Nb), the metallic *fcc*-Ir is also marked. The line corresponding to Zener index value $A_Z=1$ (isotropic Young`s modulus) is shown.

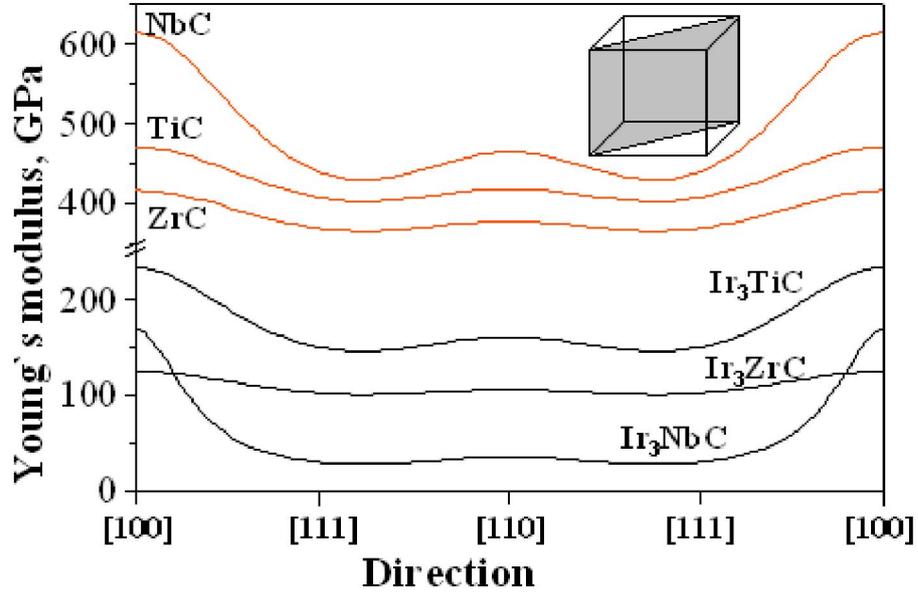

**Fig.2.** The dependence of Young`s modulus on the direction in (110) plane for Ir$_3$MC and MC (M=Ti, Zr, Nb) monocrystals.

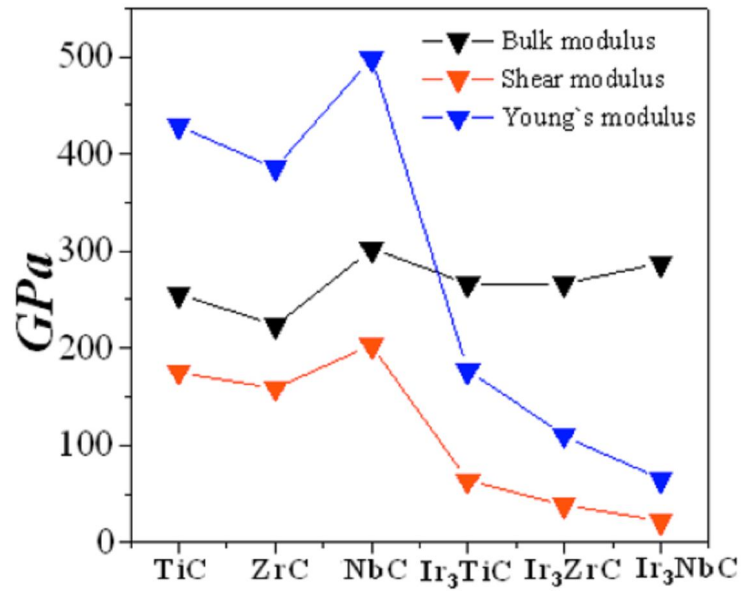

**Fig.3.** The variation of bulk, shear and Young`s moduli, as going from polycrystalline MC monocarbides to Ir$_3$MC antiperovskites.

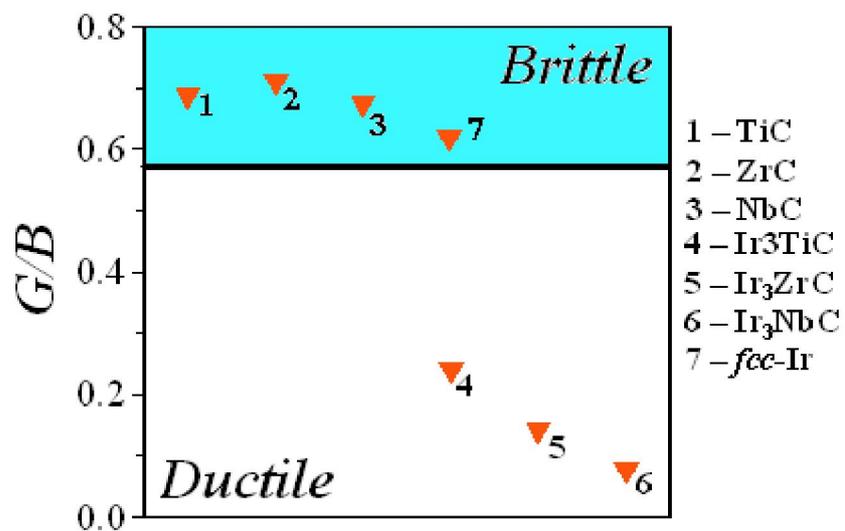

**Fig.4.** The "ductility-brittleness" diagram for Ir$_3$*M*C antiperovskites *vs* *M*C monocarbides and *fcc*-iridium. The horizontal line marks the ductility-brittleness border (*G*/*B*=0.571).

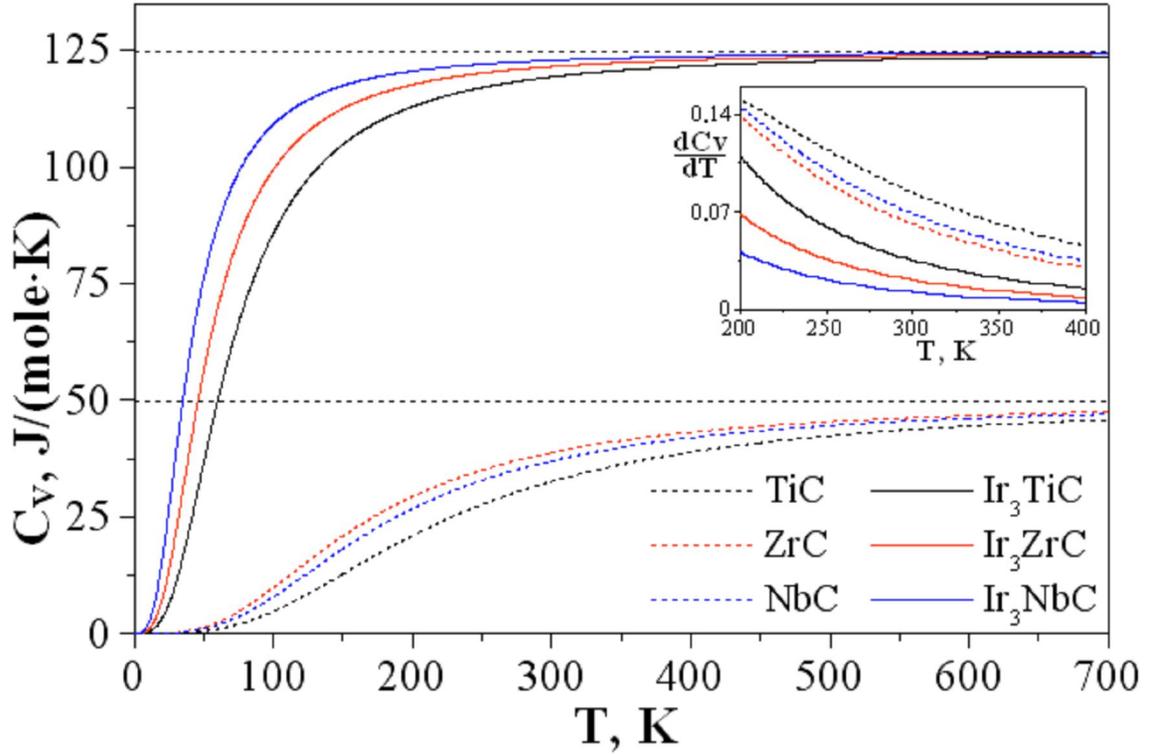

**Fig.5.** The temperature dependence of the lattice vibrational contribution to the molar heat capacity $C_V$ (within Debye model) for $Ir_3MC$ antiperovskites *vs* $MC$ monocarbides. The horizontal dotted lines specify the classical Dulong-Petit limit ($6R$ and $15R$ for monocarbides and antiperovskites, respectively). *Inset*: the derivative of $C_V$ with respect to temperature (in $J \cdot mole^{-1} \cdot K^{-2}$) in the vicinity of room temperature for all $MC$ and $Ir_3MC$ compounds is also shown.

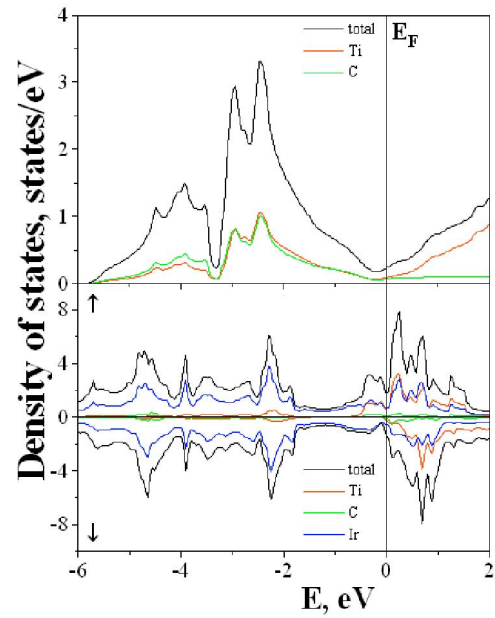

**Fig.6.** Total and partial densities of states for: TiC *vs* Ir$_3$TiC (*upper* and *lower panels*, respectively). For Ir$_3$TiC the "spin up" and "spin down" DOS components are shown individually.

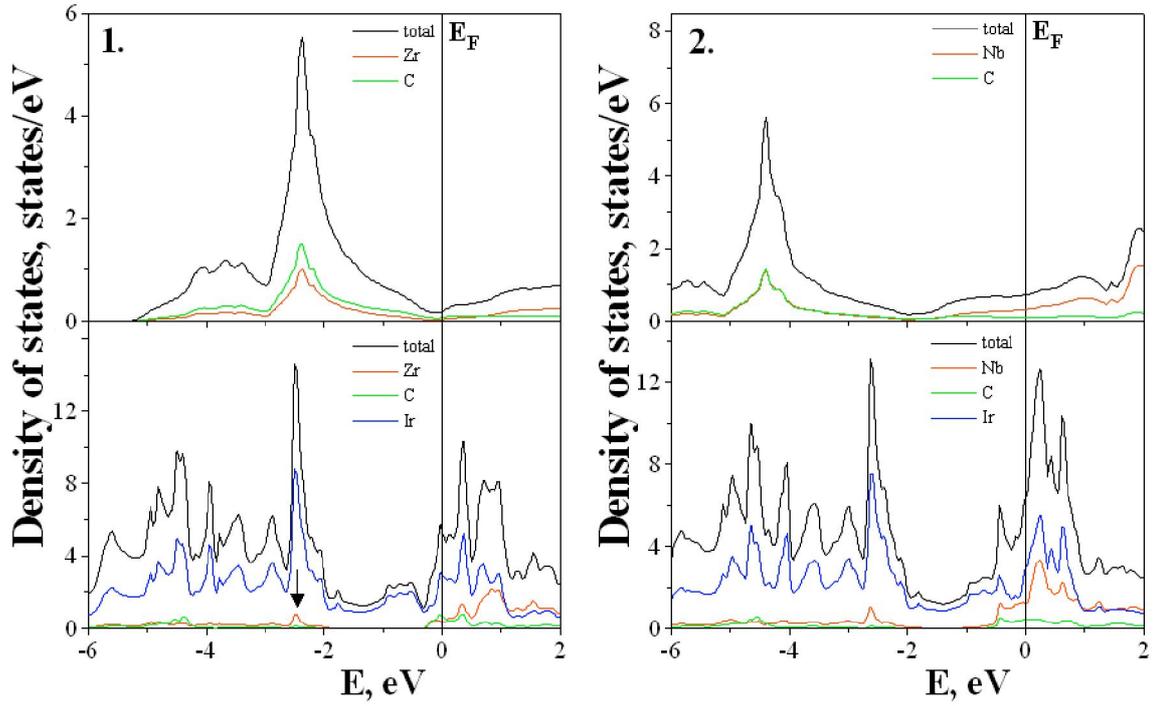

**Fig.7.** Total and partial densities of states for: (1) ZrC *vs* Ir$_3$ZrC; (2) NbC *vs* Ir$_3$NbC. For Ir$_3$ZrC – the Zr-4*d* states hybridized with Ir-5*d* states are shown by arrow.

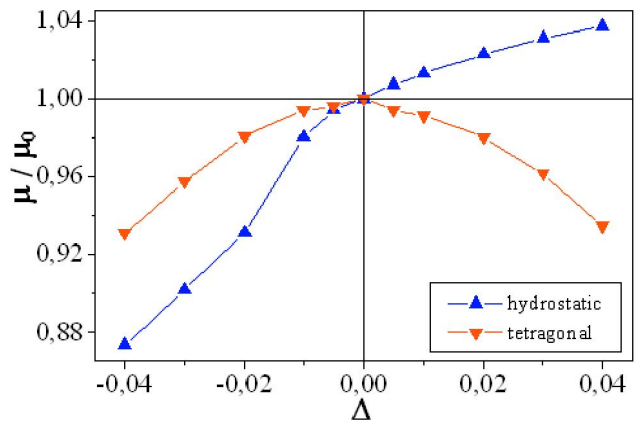

**Fig.8.** Variation of μ/μ$_0$ ratio with "hydrostatic" expansion (compression) and volume-conserving tetragonal distortion *vs* mechanical strain magnitude Δ (*see text*) in Ir$_3$TiC antiperovskite.

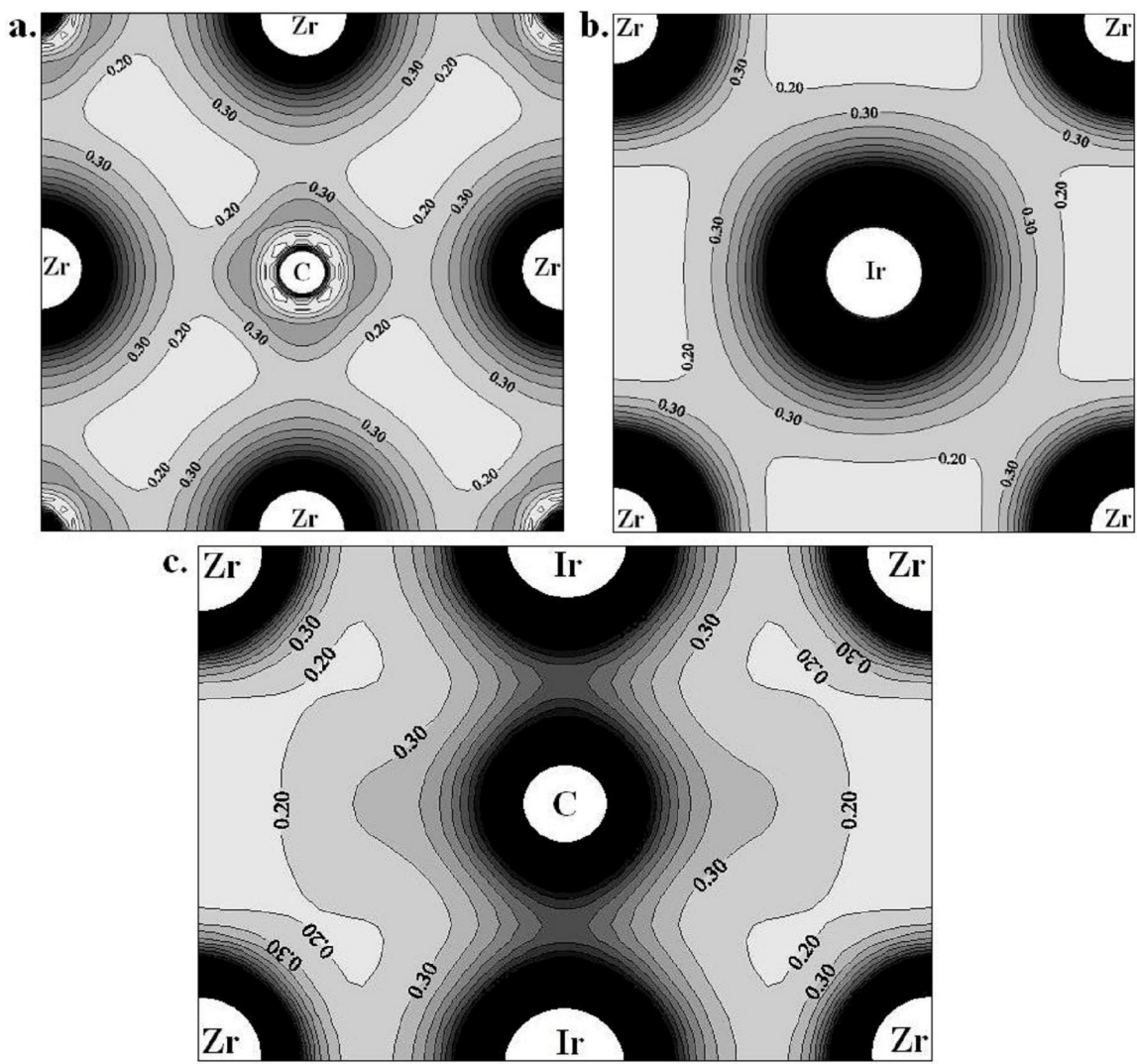

**Fig.9.** The maps of valence charge density distribution in: a. - (100) plane of ZrC binary carbide; b.,c. - (100) and (110) planes of Ir$_3$ZrC antiperovskite, respectively. The distance between contours is 0.1 $e$/Å$^3$.